\documentclass[doublecol]{epl2} 

\title{Eigenvalue and eigenspace anholonomies in hierarchical systems}
\author{Atushi Tanaka\inst{1} \and Sang Wook Kim\inst{2} 
  \and Taksu Cheon\inst{3}}
\shortauthor{A.~Tanaka, S.W.~Kim and T.~Cheon}
\institute{                    
  \inst{1} Department of Physics, Tokyo Metropolitan University,
  Hachioji, Tokyo 192-0397, Japan\\
  \inst{2} Department of Physics Education, Pusan National University,
  Busan 609-735, South Korea\\
  \inst{3} Laboratory of Physics, Kochi University of Technology,
  Tosa Yamada, Kochi 782-8502, Japan
}


\abstract{%
  An adiabatic cycle of parameters in a quantum system can yield the
quantum anholonomies, nontrivial evolution not just in phase of the
states, but also in eigenvalues and eigenstates.  Such exotic
anholonomies imply that an adiabatic cycle rearranges eigenstates even
without spectral degeneracy.  We show that an arbitrarily large
quantum circuit generated by recursive extension can also exhibit the
eigenvalue and eigenspace anholonomies.
}

\newcommand{\MYVERSION}{{\sffamily\textbf{TMU preprint} 2011-08-18}}

\makeatletter
\newcommand{\CustumHeader}[1]{
  \def\epl@stylemark{\hbox to0pt{\hskip0em \vbox to 0pt{\vss \hbox{%
          {#1} %
        }\vskip6ex}\hss}}}
\makeatother

\CustumHeader{\MYVERSION}

\usepackage{graphicx}
\usepackage{amsmath}
\usepackage{amssymb}

\newcommand{\set}[1]{\left\{ #1 \right\}}
\def\toexp{\mathop{\mathrm{exp}}}
\newcommand{\Texp}{\toexp_{\leftarrow}}
\newcommand{\AntiTexp}{\toexp_{\rightarrow}}
\newcommand{\bra}[1]{\langle{}#1{}|}
\newcommand{\ket}[1]{|{}#1{}\rangle}
\newcommand{\bracket}[2]{\langle{}#1{}|{}#2{}\rangle}
\newcommand{\ketbra}[2]{|{}#1{}\rangle\langle{}#2{}|}

\newcommand{\Cop}{\hat{C}}

\newcommand{\hc}{\text{h.c.}}
\newcommand{\AF}[1]{A^{#1}}
\newcommand{\AD}[1]{A^{\mathrm{D} #1}}

\begin{document}
\maketitle

A quantum system described by a parametric Hamiltonian goes through an
adiabatic change connecting instantaneous eigenstates when it is
subjected to a slow variation of parameters.  Berry pointed out that
this adiabatic change can yield nontrivial anholonomy in quantum phase
when the trajectory of the parameter variation is closed to form a
loop, for which he coined the term 
geometric phase~\cite{Berry-PRSLA-392-45}.

\begin{figure}[t]
  \centering
  \includegraphics[width=4cm]{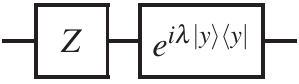}
  \caption{
    A simple quantum circuit that exhibits eigenvalue and eigenspace 
    anholonomies,
      where $Z=\ketbra{0}{0}-\ketbra{1}{1}$ and
      $\ket{y} \equiv (\ket{0}-i\ket{1})/{\sqrt{2}}$.
      Note that $e^{i {\lambda}\ketbra{y}{y}}$ is a phase shift gate:
      If the input is $\ket{y}$, this gate put a phase factor $e^{i\lambda}$
      to the qubit.
      On the other hand, if the input state is orthogonal to $\ket{y}$,
      this gate do not change the state of the qubit.
    This system can be interpreted as 
    a kicked spin system~\cite{Tanaka-PRL-98-160407}. 
    Although the eigenvalue and eigenspace anholonomies may be 
    considered to be rather exotic,
    these anholonomies appear even in such a simple system.
    We will examine its $N$-qubit extensions in the main text.
  }
  \label{fig:us}
\end{figure}

Anholonomy under cyclic adiabatic parameter variation need not be
limited to the phase of an eigenstate, but can involve
the interchange of eigenvalues and eigenstates,
since what is
required 
after the return to the original parameter value
is the identity of the whole set
of eigensystem.  Curiously, however, this possibility has been
overlooked until recently, when examples of quantum system exhibiting
eigenvalue and eigenspace anholonomy have been found, first in the
Hamiltonian spectra of one-dimensional system with generalized point
interaction~\cite{Cheon-PLA-248-285}, and then in the Floquet spectra 
of time-periodic kicked-spin~\cite{Tanaka-PRL-98-160407} 
(see also, Fig.~\ref{fig:us}). 
In hindsight, quantum anholonomy of Wilczek and Zee, in
which eigenstates belonging to a single degenerate eigenvalue undergo
mutual exchange and mixing~\cite{Wilczek-PRL-52-2111}, 
can be thought of as a precursor to this new type.
Since then, further examples
of novel type of quantum anholonomy have been found both in
Hamiltonian~\cite{Cheon-PLA-374-144,Tanaka-PRA-82-022104} and 
Floquet systems~\cite{Miyamoto-PRA-76-042115,Cheon-EPL-85-20001}. 
In one example, even the requirement
of adiabaticity has been lifted, and the new type of anholonomy is
shown to persist in a system with non-adiabatic cyclic parameter
variation~\cite{Tanaka-PRA-82-022104}.

Lately, it has been shown that all quantum holonomy, 
namely, 
Berry,
Wilczek-Zee and the novel type having the 
eigenvalue exchange
can be
described by a unified formulation which is 
built 
upon 
the generalized
Mead-Truhlar-Berry gauge connection~\cite{Cheon-EPL-85-20001}. 
The structure of gauge connection in the
anholonomy of the new type has been examined in terms of the theory of
Abelian gerbes~\cite{Viennot-JPA-42-395302}.  
It is also clarified that the exceptional point, the
singularity of gauge connection in complex plane, plays a crucial role
in the new type of quantum holonomy~\cite{Kim-PLA-374-1958}.
Applications to adiabatic manipulations of quantum states, including 
quantum computation~\cite{NielsenChuang}, 
are promising~\cite{Tanaka-PRL-98-160407,Tanaka-PRA-81-022320},
since the eigenvalue and eigenspace anholonomies are stable 
against perturbations that retains the periodicity of the parameter 
space~\cite{Tanaka-PRL-98-160407,Miyamoto-PRA-76-042115}.

Until now, there has been no known composite system that exhibits the
new type of quantum anholonomy. In other words, all the conventional
examples are ``one-body'' type systems. This raises the question on whether
there is any way to realize the eigenvalue and eigenspace anholonomies
in quantum composite systems. 
This question is not only
fundamental for the anholonomies, but also important for the
application to quantum computation~\cite{Tanaka-PRA-81-022320}: 
It is essential to deal with more than one qubit and to find a systematic
way to generate multi-qubit systems starting from a single qubit.

In this letter, we propose a systematic way to construct quantum circuits
that exhibit the eigenvalue and eigenspace anholonomies based upon multiple
qubits. 
The obtained multi-qubit systems clearly show that the hierarchical structure
exists in the resultant quantum circuits.
We also
examine the influence of the hierarchical structure on the non-Abelian
gauge connection associated with the eigenspace anholonomy.

\section{Preliminaries --- a one-body example}
\label{sec:onebody}
We first explain the constituent 
building block 
of our many-body examples.
This is a quantum circuit on a qubit.
The eigenvalue and eigenspace anholonomies in the constituent are
also explained using a gauge theoretical approach 
for the anholonomies~\cite{Cheon-EPL-85-20001}.

We introduce our 
simplest 
example (Fig.~\ref{fig:us}):
\begin{align}
  \hat{u}(\lambda)
  \label{eq:u}
  = e^{i\lambda \ketbra{y}{y}} \hat{Z}
  ,
\end{align}
where 
$\hat{X}\equiv \ketbra{0}{1} + \hc$,
$\hat{Y}\equiv i\ketbra{1}{0}+\hc$,
and 
$\hat{Z}\equiv \ketbra{0}{0} - \ketbra{1}{1}$.
$\ket{y} \equiv (\ket{0}-i\ket{1})/{\sqrt{2}}$
satisfies $\hat{Y}\ket{y}=-\ket{y}$.
The first factor in Eq.~\eqref{eq:u} is the phase shift gate 
where the $y$-axis of the control qubit is chosen.
We solve the eigenvalue problem of $\hat{u}(\lambda)$. 
Let $z(n; \lambda)$ denote the $n$-th eigenvalue of 
the unitary operator $\hat{u}(\lambda)$ ($n=0,1$). 
Since $z(n; \lambda)$ is a a unimodular complex number,
we introduce a real number $\theta(n; \lambda)$ that 
satisfies
$z(n; \lambda) = \exp\{i\theta(n; \lambda)\}$:
\begin{align}
  \label{eq:theta-one-body}
  \theta(n; \lambda) = n\pi + \frac{\lambda}{2}
  ,
\end{align}
which is called 
an
eigenangle.
The corresponding eigenvectors are
\begin{equation}
  \label{eq:EigenvectorOfU}
  \begin{split}
  \ket{0; \lambda}&
  \equiv \cos\frac{\lambda}{4}\ket{0}+\sin\frac{\lambda}{4}\ket{1}
  ,\\
  \ket{1; \lambda}&
  \equiv \cos\frac{\lambda}{4}\ket{1}-\sin\frac{\lambda}{4}\ket{0}
  ,
  \end{split}
\end{equation}
where the phases of the eigenvectors are chosen so as to simplify
the following analysis.

$\hat{u}(\lambda)$ is periodic with $\lambda$,
namely $\hat{u}(\lambda+2\pi)=\hat{u}(\lambda)$.
Let $C$ denote the closed path of $\lambda$ from $\lambda=0$
to $2\pi$. 
Both the spectral set $\set{e^{i\theta(n; \lambda)}}_{n=0,1}$ and 
the set of projectors
$\set{\ketbra{n; \lambda}{n; \lambda}}_{n=0,1}$ 
obey 
the same 
periodicity of {$\hat{u}(\lambda)$}.
However, each eigenvalue $e^{i\theta(n; \lambda)}$ and 
each projector $\ketbra{n; \lambda}{n; \lambda}$ have a longer period.
Namely, $\hat{u}(\lambda)$ exhibits eigenvalue and eigenspace anholonomies
under adiabatic parametric change along the closed path $C$,
where 
eigenvalues and eigenvectors are respectively exchanged.

We outline a theoretical framework for the eigenangle and eigenspace
anholonomies using the one-body example $\hat{u}(\lambda)$.
First, we examine the anholonomy in eigenangles. Using integers
$s(n)$ and $r(n)$, the parametric dependence of eigenangle 
is arranged as
\begin{equation}
  \theta(n; \lambda + 2\pi)
  = \theta(s(n); \lambda) + 2\pi r(n)
  .
\end{equation}
Namely, the $n$-th eigenangle arrives
the $s(n)$-th eigenangle after a cycle $C$.
On the other hand, $r(n)$ is ``a winding number'' of quasienergy
(cf. It is shown that $\sum_n r(n)$ offers a topological character
for the ``Floquet operator'' 
$\hat{u}(\lambda)$~\cite{Kitagawa-PRB-82-235114}).
We introduce a $2\times 2$ matrix $S(C)$ whose elements are defined as
\begin{align}
  \{S(C)\}_{n',n}\equiv \delta_{n',s(n)}
  .
\end{align}
From Eq.~\eqref{eq:theta-one-body}, we have
$s(n) = \overline{n}$ and $r(n) = n$,
where $\overline{0}=1$ {and} $\overline{1}=0$.
Accordingly, the permutation matrix
$\{S(C)\}_{n',n} = \delta_{n',\bar{n}}$
describes a cycle whose length is $2$.

Second, we examine the geometric phase factors associated with 
the adiabatic time evolution along the cycle $C$. 
We introduce a holonomy matrix $M(C)$~\cite{Cheon-EPL-85-20001},
whose $(n',n)$-th elements is the overlapping integral 
$\bracket{n'; \lambda}{n; \lambda+2\pi}$, 
where $\ket{n;\lambda}$ is supposed to satisfy
the parallel transport condition~\cite{Stone-PRSLA-351-141} along $C$
for the non-degenerate eigenspace,
i.e., {$\bra{n;\lambda}\partial_{\lambda}\ket{n; \lambda}=0$}.
Due to the periodicity of $\hat{u}(\lambda)$, 
$\ket{n; \lambda+2\pi}$ is either parallel or perpendicular to 
$\ket{n; \lambda}$.
In the former case, the diagonal elements of $M(C)$ are
the Berry's geometric phase factors~\cite{Berry-PRSLA-392-45}.
The latter case implies
the presence of the eigenvalue and eigenspace anholonomies and 
the off-diagonal
elements provides Manini-Pistolesi's gauge 
invariants~\cite{Manini-PRL-85-3067}. 
It is worth to remark that the nodal-free geometric phase factors, which are
the eigenvalues of $M(C)$,
offers the geometric phases for both cases~\cite{Ericsson-PLA-372-596}.
An extension of Fujikawa formalism 
for the geometric phase~\cite{Fujikawa-PRD-72-025009}
offers a gauge covariant expression of $M(C)$~\cite{Cheon-EPL-85-20001}:
\begin{align}
  \label{eq:Mcov}
  M(C)
  =
  \AntiTexp\left(-i\int_{C} A(\lambda)\upd\lambda\right)
  \Texp\left(i\int_{C} \AD{}(\lambda)\upd\lambda\right)
  ,
\end{align}
where $\Texp$ and $\AntiTexp$ are path-ordered and anti-path-ordered
exponentials, respectively.
Gauge connections 
$\{A(\lambda)\}_{n',n}
\equiv i\bra{n';\lambda}\partial_{\lambda}\ket{n;\lambda}$ and 
{$\AD{}(\lambda)
\equiv \delta_{n',n}\{A(\lambda)\}_{n',n}$} 
are also introduced.
The second factor in the right hand side of Eq.~\eqref{eq:Mcov}
describes time evolution in terms of 
adiabatic
basis vectors
$\ket{n; \lambda(t)}$~\cite{Anandan-PLA-133-171} (see also, 
Eq.~(3) in Ref.~\cite{Cheon-EPL-85-20001}). 
On the other hand, the first factor in Eq.~\eqref{eq:Mcov}
is introduced so as to incorporate the multiple-valuedness
of $\ket{n;\lambda}$, which is nothing but the eigenspace 
anholonomy~\cite{Cheon-EPL-85-20001}.
The gauge connection in our model~(\ref{eq:u}),
under the gauge specified by Eq.~\eqref{eq:EigenvectorOfU},  
is
\begin{align}
  A(\lambda) = \frac{1}{4}Y
  ,\quad\text{where}\quad
  \begin{bmatrix}
    Y_{00}&Y_{01}\\ Y_{10}&Y_{11}
  \end{bmatrix}
  = 
  \begin{bmatrix}
    0&-i\\ i&0
  \end{bmatrix}
  ,
\end{align}
which implies $\AD{}(\lambda)=0$.
Accordingly, we obtain
$M(C) = -iY$, or, equivalently
\begin{align}
  \{M(C)\}_{n',n}
  = \{S(C)\}_{n',n}(-1)^{n}
  .
\end{align}
Namely, $M(C)$ is composed by two parts, the permutation matrix $S(C)$ and 
the phase factors $(-1)^{n}$.

\section{Recursive construction of $N$-qubit circuit}

A crucial ingredient of our $N$-body extension of $\hat{u}(\lambda)$
is the following ``super-operator'' $\hat{D}[\cdot]$ 
\begin{align}
  \label{eq:Ddef}
  \hat{D}[\hat{U}]\equiv \Cop^y[\hat{U}] (\hat{Z}\otimes\hat{1})
\end{align}
in which
\begin{align}
  \label{eq:CopDef}
  \Cop^y[U]
  \equiv (\hat{1}-\ketbra{y}{y})\otimes \hat{1}
  + \ketbra{y}{y}\otimes \hat{U}
\end{align}
is a controlled-unitary gate, where the ``axis'' of the 
control-bit is chosen to be in the ``$y$-direction''.

In order to expose the one body quantum circuit $\hat{u}(\lambda)$ 
hidden in $\hat{D}[\cdot]$, 
we examine $\hat{D}[\cdot]$ with the global phase
gate $e^{i\lambda}\hat{1}_{\mathrm{A}}$ for an ancilla,
where $\hat{1}_{\mathrm{A}}$ is the identity operator of the ancilla:
$\hat{D}[e^{i\lambda}\hat{1}_{\mathrm{A}}]
= \hat{u}(\lambda)\otimes\hat{1}_{\mathrm{A}}.$
Namely, we may say that $\hat{u}(\lambda)$ is an extension of
the global phase gate on an ancilla with $\hat{D}[\cdot]$.
This interpretation suggests the following $N$-body extension.

\begin{figure}[t]
  \centering
  \includegraphics[width=8cm]{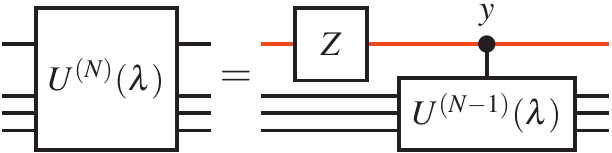}
  \caption{(Color online) 
    Recursive construction of $\hat{U}^{(N)}(\lambda)$. 
    The label $y$ over the control qubit
    indicates the ``axis'' for the control.
    $\hat{U}^{(N)}(\lambda)$ is made of 
    a qubit, indicated by the uppermost line (colored),
    and $N-1$ qubits, indicated by the lower lines.
  }
  \label{fig:Urecur}
\end{figure}

\begin{figure}[t]
  \centering
  \includegraphics[width=6cm]{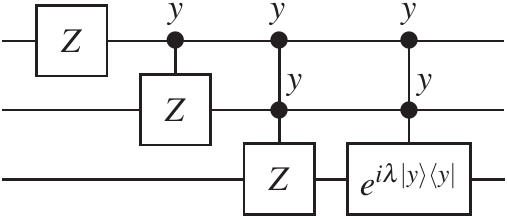}
  \caption{
    Circuit for $\hat{U}^{(3)}(\lambda)$.
    For notation, see, Fig.~\ref{fig:Urecur}.
    The corresponding algebraic expression of the circuit is
    $\hat{U}^{(3)}(\lambda)
    = \{(\Cop^y)^2[e^{i\lambda\ketbra{y}{y}}]\}
    \{(\Cop^y)^2[Z]\}\{\Cop^y[Z]\}(Z\otimes1\otimes1),$
    where  $(\Cop^y)^n[U] = \Cop^y[(\Cop^y)^{n-1}[U]]$ for $n>1$
    and $(\Cop^y)^1[U] = \Cop^y[U]$.
    It is straightforward to see that 
    $\hat{U}^{(N)}(\lambda)$ is composed by
    $Z$, $\Cop^y[Z]$, $\dots$, $(\Cop^y)^{N-1}[Z]$ 
    and $(\Cop^y)^{N-1}[e^{i\lambda\ketbra{y}{y}}]$.
    Because $(\Cop^y)^{n}[Z]$ can be composed by
    $\mathcal{O}(n^2)$ quantum gates 
    (see, e.g, \S 4.3 in Ref.~\cite{NielsenChuang}),
    $\hat{U}^{(N)}(\lambda)$ can be efficiently implemented
    by quantum gates.
  }
  \label{fig:u3}
\end{figure}

A family of quantum circuits 
$\hat{U}^{(N)}(\lambda)$ on $N$-qubits is recursively defined
in the following.
For $N=1$, we set $\hat{U}^{(1)}(\lambda) = \hat{u}(\lambda)$,
which is examined above.
For $N>1$, we compose $\hat{U}^{(N)}(\lambda)$
from a $(N-1)$-qubit circuit $\hat{U}^{(N-1)}(\lambda)$,
adding a qubit (Fig.~\ref{fig:Urecur}):
\begin{align}
  \label{eq:UNrec}
  \hat{U}^{(N)}(\lambda)
  \equiv \hat{D}[\hat{U}^{(N-1)}(\lambda)]
  .
\end{align}
We depict the quantum circuit with $N=3$ in Fig.~\ref{fig:u3}.
It is worth pointing out the scalability of $\hat{U}^{(N)}(\lambda)$,
i.e., only ${\rm poly}(N)$ quantum gates are required to 
construct $\hat{U}^{(N)}(\lambda)$.

\newcommand{\tmpTh}{\Theta}

We will solve the eigenvalue problem of $\hat{U}^{(N)}(\lambda)$.
A complete set of quantum numbers for $\hat{U}^{(N)}(\lambda)$
is $(n_N, n_{N-1}, \cdots, n_1)$, where $n_j\in \set{0,1}$.
Let $\ket{n_N, n_{N-1}, \cdots, n_1; \lambda}$ 
and $\theta^{(N)}(n_N, n_{N-1}, \cdots, n_1; \lambda)$ 
denote an eigenvector and the corresponding eigenangle
of $\hat{U}^{(N)}(\lambda)$,
respectively.
We will
obtain recursion relations for eigenangles and eigenvectors.
Suppose that we have an eigenvector 
$\ket{\Psi}$ 
and 
the corresponding eigenangle
$\tmpTh$ 
of the smaller quantum circuit 
$\hat{U}^{(N-1)}(\lambda)$.
From Eq.~\eqref{eq:UNrec}, 
$\ket{n_N; \tmpTh}\otimes\ket{\Psi}$
is an eigenvector of $\hat{U}^{(N)}(\lambda)$, where
$n_N\in\set{0,1}$.
The corresponding eigenangle is 
$\theta^{(1)}(n_N; \tmpTh)$.
Accordingly 
the recursion relations are
\begin{align}
  \label{eq:EigAngleRecursion}
  &
  \theta^{(N)}(n_{N}, n_{N-1}, \dots, n_1;\lambda)
  \nonumber\\ &
  = 
    \theta^{(1)}%
  (n_{N}; \theta^{(N-1)}(n_{N-1}, \dots, n_1;\lambda))
\intertext{and}
&
  \label{eq:EigVectorRecursion}
  \ket{n_{N}, n_{N-1},\dots, n_1;\lambda}
  \nonumber\\ &
  = 
  \ket{n_{N}; \theta^{(N-1)}(n_{N-1}, \dots, n_1;\lambda)}
  \nonumber\\ &\qquad
  \otimes\ket{n_{N-1},\dots, n_1;\lambda}
  .
\end{align}
Hence we obtain 
the
eigenangle
\begin{align}
  \label{eq:theta_by_mN}
  &
  \theta^{(N)}(n_{N}, \dots, n_1;\lambda)
  \nonumber\\ &
  = \frac{2\pi}{2^N}\left\{m_N(n_{N}, n_{N-1}, \dots, n_1) 
    + \frac{\lambda}{2\pi}\right\}
  ,
\end{align}
where 
\begin{align}
  \label{eq:mN}
  m_N(n_{N}, \dots, n_1) 
  &
  \equiv \sum_{j=1}^{N} 2^{j-1} n_j
  .
\end{align}
This is 
called a 
principal quantum number.
Note that
$\hat{U}^{(N)}(\lambda)$ has no spectral degeneracy.
Eq.~\eqref{eq:mN} shows
$n_{N}, \dots, n_1$ are the coefficients
of the binary expansion of $m_N$.
Because of the simplicity of the correspondence between
$m_N$ and $(n_{N}, \dots, n_1)$, we will identify them in the following.
An eigenvector of $\hat{U}^{(N)}(\lambda)$ is
\begin{align}
  &
  \ket{n_{N},\dots, n_1;\lambda}
  \nonumber\\ &
  = 
  |n_N ; \theta^{(N-1)}(m_{N-1};\lambda)\rangle
  \otimes\dots
  \nonumber\\ &\qquad
  \otimes|n_{2}; \theta^{(1)}(m_{1};\lambda)\rangle
  \otimes|n_{1}; \lambda\rangle
  .
\end{align}

\section{Analysis of exotic anholonomies}

We examine the parametric 
dependences
of eigenangles and eigenprojectors
of $\hat{U}^{(N)}(\lambda)$, along the cycle $C$, i.e., 
$\lambda\mapsto\lambda+2\pi$.
Let $(n_N,\dots, n_1)$, or equivalently, $m_N$, 
be the set of quantum numbers of
the initial eigenstate.
After the completion of 
the parametric variation along
the cycle $C$, 
the quantum numbers are rearranged, which shows the anholonomies take place.
Let $s^{(N)}_j(m_N; C)$ ($\in\set{0,1}$) denote the value of 
the
$j$-th 
quantum number ($1\le j \le N$).
We introduce a $2^N\times2^N$ permutation matrix $S^{(N)}(C)$ whose
elements are determined by $s^{(N)}_j(m_N; C)$:
\begin{align}
  \label{eq:SNdef}
  \{S^{(N)}(C)\}_{m'_N, m_N}
  = \prod_{j=1}^N \delta_{n'_j, s_j(m_N; C)}
  ,
\end{align}
which represents an 
anholonomy in quantum numbers.

Now we show that
$S^{(N)}(C)$ 
can be obtained
from 
the parametric
dependence
of eigenangles:
\begin{align}
  \label{eq:srBalance}
  &
  \theta^{(N)}(m_N;\lambda+2\pi)
  \nonumber\\&
  = \theta^{(N)}(s^{(N)}(m_N; C);\lambda)
  + 2\pi r^{(N)}(m_N; C)
  ,
\end{align}
where we abbreviate 
the collection of the quantum numbers
$s^{(N)}_N(m_N; C), \dots, s^{(N)}_1(m_N; C)$
as $s^{(N)}(m_N; C)$.
An integer $r^{(N)}(m_N; C)$ is a ``winding number'' of 
$\theta^{(N)}(m_N;\lambda)$ in the periodic space of eigenangle.

In our model (Eq.~(\ref{eq:UNrec})), 
$s^{(N)}_N$ and $r^{(N)}$ can be obtained through recursion relations.
Here we show only the solutions:
\begin{align}
  \label{eq:srForRank1}
  s^{(N)}_N(n_{N}, \dots, n_1)
  &
  = 
  \begin{cases}
    \overline{n_{N}} & \text{for $n_{N-1}\cdots n_1 = 1$}\\
    n_{N} & \text{otherwise}\\
  \end{cases}
  ,
  \intertext{and}
  r^{(N)}(n_{N}, \dots, n_1)
  &
  = 
  n_N\cdots n_1
  ,
\end{align}
for $N>1$, and 
$s^{(1)}_1(n_1) = \overline{n_1}$ and $r^{(1)}(n_1) = n_1$.
Hence the permutation matrix $S^{(N)}$ 
contains only a cycle
whose length is
$2^N$.
The itinerary of $n_N,\dots,n_1$ with $N=3$, for example,
is the following
\begin{align}
  &
  000 \mapsto 
  001 \mapsto 
  010 \mapsto 
  011 
  \nonumber\\
  \mapsto &
  100 \mapsto 
  101 \mapsto 
  110 \mapsto 
  111 \mapsto 
  000
  .
\end{align}
In terms of the principal quantum number $m_N$, this itinerary
can be described in a simple way:
When $m_N < 2^N-1$, 
at every parametric variation of $\lambda$ by $2\pi$,
$m_N$ increases by unity.
Otherwise, $m_N$ 
becomes
zero.
Hence, 
after encircling the path $C$ by $2^N$ times,
$m_N$ reruns to 
the initial point.

\section{Adiabatic geometric phase}
We examine the holonomy matrix
\begin{align}
  \label{eq:MNintro}
  \{M^{(N)}(C)\}_{m'_N,m_N}
  = \bracket{m'_N;\lambda}{\psi^{(C)}(m_N; \lambda)}
  ,
\end{align}
where $\ket{\psi^{(C)}(m_N; \lambda)}$ is obtained
by the parallel transport of $\ket{m_N; \lambda}$
along the path $C$.
$M^{(N)}(C)$ incorporates two aspects of adiabatic cycle along $C$.
One is the change in the quantum numbers, which is described
by $S^{(N)}(C)$ (Eq.~\eqref{eq:SNdef}).
The other involves the geometric phase, in a generalized 
sense~\cite{Manini-PRL-85-3067,Samuel-PRL-60-2339}. Let $\sigma^{(N)}(m_N)$ be
the phase factor associated with the eigenspace initially labeled by $m_N$.
These two factors comprise the holonomy matrix
$\{M^{(N)}(C)\}_{m'_N, m_N} 
= \{S^{(N)}(C)\}_{m'_N, m_N}\sigma^{(N)}(m_N; C).$
We will obtain the phase factor
$\sigma^{(N)}(m_N; C)$
using a gauge covariant expression of $M^{(N)}(C)$ (Eq.~\eqref{eq:Mcov}).
Here the non-Abelian gauge connection is
\begin{align}
  &
  \{A^{(N)}(\lambda)\}_{m'_N,m_N}
  \equiv 
  \bra{m'_N;\lambda}
  \left[i\partial_{\lambda}\ket{m_N;\lambda}\right]
  .
\end{align}
Because of the absence of the spectral degeneracy in 
$\hat{U}^{(N)}(\lambda)$, the 
diagonal
part of
$A^{(N)}(\lambda)$ is 
$
\{\AD{(N)}(\lambda)\}_{m'_N,m_N}
\equiv 
\delta_{m'_N,m_N}
\{A^{(N)}(\lambda)\}_{m_N,m_N}
$.

We will obtain $A^{(N)}(\lambda)$ through recursion relations.
Note that the results for the case $N=1$ are already obtained above.
For $N>1$, 
the Leibniz rule in the derivative of Eq.~(\ref{eq:EigVectorRecursion})
suggests the decomposition of the gauge connection
$\AF{(N)}(\lambda)
= \AF{(N)}_{\mathrm{H}}(\lambda)+\AF{(N)}_{\mathrm{L}}(\lambda)$, where
\begin{align}
  &
  \{\AF{(N)}_{\mathrm{H}}(\lambda)\}_{m'_N,m_N}
  \nonumber\\&
  \equiv
  \left\{%
      A(\theta^{(N-1)}(m_{N-1}';\lambda))
    \right\}_{n_N',n_N}
  \nonumber\\ &\qquad{}\times
  \left\{\partial_{\lambda}\theta^{(N-1)}(m_{N-1}';\lambda)\right\}
  \prod_{j=1}^{N-1}\delta_{n_{j}',n_{j}}
  \\
  &
  \{\AF{(N)}_{\mathrm{L}}(\lambda)\}_{m'_N,m_N}
  \nonumber\\&
  \equiv
  \bracket{n'_N (\theta^{(N-1)}(m_{N-1}';\lambda))}%
  {n_N (\theta^{(N-1)}(m_{N-1};\lambda))}
  \nonumber\\ &\qquad{}\times
  \left\{\AF{(N-1)}(\lambda)\right\}_{m'_{N-1},m_{N-1}}
  .
\end{align}
We also have a similar recursion relation for ``diagonal'' gauge connection 
$\AD{(N)}(\lambda)$.
As we have already chosen the gauge that satisfy
$\AD{}(\lambda)=0$ in the one-body problem,
we obtain 
$\AD{(N)}(\lambda)=0$ for all $N$ from the recursion relations.
It is straightforward to obtain
\begin{align}
  \AF{(N)}_{\mathrm{L}}(\lambda)
  &
  = 
  e^{i\pi 2^{-N}Y\otimes J_{\mathrm{D}}^{(N-1)}}
  \left\{1\otimes A^{(N-1)}(\lambda)\right\}
  \nonumber\\&\qquad\times
  e^{-i\pi 2^{-N} Y\otimes J_{\mathrm{D}}^{(N-1)}}
  ,
  \intertext{and}
  \AF{(N)}_{\mathrm{H}}(\lambda)
  &
  = \frac{1}{2^{N+1}}Y\otimes1^{(N-1)}
  ,
\end{align}
where
$
\left\{J_{\mathrm{D}}^{(N)}\right\}_{m'_{N},m_{N}}
\equiv m_{N}\delta_{m'_N,m_N}.$
Because $\AF{(N)}_{\mathrm{H}}(\lambda)$ is independent of $\lambda$,
$\AF{(N)}_{\mathrm{L}}(\lambda)$ as well as $A^{(N)}(\lambda)$ are
also independent of $\lambda$.
Hence, in our model, $A^{(N)}(\lambda)$ is independent 
of $\lambda$. Furthermore, 
$\AF{(N)}_{\mathrm{H}}(\lambda)$ commutes with $\AF{(N)}_{\mathrm{L}}(\lambda)$.
Now it is straightforward to obtain 
%
\begin{align}
  \left\{M^{(N)}(C)\right\}_{m'_N, m_N}
  &
  = 
  \{S^{(N)}(C)\}_{m'_N, m_N}
  (-1)^{r^{(N)}(m_N)}
  \nonumber\\&\qquad\times
  \sigma^{(N-1)}(m_{N-1}; C)
  ,
\end{align}
which implies a recursion relation for $\sigma^{(N)}$:
\begin{align}
  \sigma^{(N)}(m_{N}; C)
  =
  (-1)^{r^{(N)}(m_N)}
  \sigma^{(N-1)}(m_{N-1}; C)
  .
\end{align}

\section{Manini-Pistolesi gauge invariant}
The holonomy matrix $M^{(N)}(C)$ is a gauge covariant quantity, from
which we can construct a {\it gauge invariant} Manini-Pistolesi 
phase factor~\cite{Manini-PRL-85-3067}
\begin{align}
  \label{eq:gammaMPN}
  \gamma_{\mathrm{MP}}^{(N)}(C) 
  = 
  \prod_{n_{N}=0}^{1}\cdots\prod_{n_1=0}^{1} 
  \sigma^{(N)}(n_{N},\dots, n_1; C),
\end{align}
which turns out to be a sole 
nontrivial 
invariant phase factor
of the system.
We explain how Eq.~(\ref{eq:gammaMPN}) is obtained
through the $2^N$ repetitions of the adiabatic cycle $C$.
Assume we start from an eigenstate specified by a set of 
quantum number $(n_{N},\dots, n_1)$. 
Let $(n_{N}(j),\dots, n_{1}(j))$ denote the value of
quantum numbers after the completion of $j$-th cycle.
At $j=2^N$, $(n_{N}(j),\dots, n_{1}(j))$ returns to
the initial point. From $M^{(N)}(C)$, 
$\gamma_{\mathrm{MP}}^{(N)}(C)$ is defined as
\begin{align}
  &
  \gamma_{\mathrm{MP}}^{(N)}(C)
  \nonumber\\&
  = \prod_{j=1}^{2^{N}-1}
  \{M(C)\}_{(n_{N}(j+1),\dots, n_{1}(j+1)), (n_{N}(j),\dots, n_{1}(j))}
  .
\end{align}
Because $(n_{N}(j),\dots, n_{1}(j))$ experiences all the combinations
of $\set{0,1}^N$, 
we have
$\gamma_{\mathrm{MP}}^{(N)}(C)
  = \prod_{n_{N}=0}^{1}\cdots\prod_{n_1=0}^{1} 
  \{M(C)\}_{s(n_{N},\dots, n_{1}), (n_{N},\dots, n_{1})},$
which implies Eq.~(\ref{eq:gammaMPN}).
The meaning of $\gamma_{\mathrm{MP}}^{(N)}(C)$ is straightforward; it is
the Berry phase obtained after $2^N$ repetitions of the loop $C$
in the parameter space.  In our model, we have
$$
 \gamma_{\mathrm{MP}}^{(N)}(C) = -1
$$ 
for arbitrary $N > 0$, which suggests that the anholonomy found in the
model is a ``halfway evolution'' 
to
Longuet-Higgins 
anholonomy~\cite{Herzberg-DFS-35-77}.

\section{Summary and Discussion}
We have introduced a family of a multi-qubit systems that display 
the eigenvalue and the eigenspace anholonomies. The systems considered 
here can be regarded as a quantum map under a rank-one 
perturbation~\cite{Combesqure-JSP-59-679,Miyamoto-PRA-76-042115}.
As this family is
composed in a recursive way, their eigenvalues and eigenvectors 
exhibit a hierarchical structure. Furthermore, 
since
these examples have explicit 
analytic
expressions of eigenvalues and 
eigenvectors, we have examined the resultant eigenspace anholonomy 
using the extended Fujikawa formalism. The structure of the non-Abelian 
gauge connection also reflects the recursive construction. 
The path-ordered exponential of the gauge connection has an explicit 
analytical expression, which allows us to examine the holonomy matrix 
throughout.

Our quantum circuit $\hat{U}^{(N)}(\lambda)$ can be utilized as
a reference to construct another quantum circuits that retains 
the eigenvalue and the eigenspace anholonomies. More precisely, 
while we vary $\hat{U}^{(N)}(\lambda)$ keeping both the periodicity 
in $\lambda$ and the unitarity, 
the anholonomies survive
until we encounter a spectral 
degeneracy~\cite{Tanaka-PRL-98-160407,Miyamoto-PRA-76-042115}. 
Because $\hat{U}^{(N)}(\lambda)$ has no spectral degeneracy as shown 
in Eq.~\eqref{eq:theta_by_mN}, there are a lot of quantum circuits 
that exhibit the anholonomies around $\hat{U}^{(N)}(\lambda)$.

On the other hand, the eigenspace and eigenvalue anholonomies are 
generally fragile against the increment of the degrees of freedom. 
For example, when two qubits each of which exhibits the anholonomies 
are composed without any interaction,
i.e., $\hat{u}(\lambda)\otimes\hat{1} + \hat{1} \otimes \hat{u}(\lambda)$,
the anholonomies do not survive in the resultant composite system.
Hence the realization of eigenspace and eigenvalue anholonomies
in a quantum composite system is not straightforward. 
This 
also explains why the eigenspace and eigenvalue anholonomies are 
rather 
uncommon.

Nevertheless, our recursive construction offers a way to realize 
the anholonomies in quantum 
composite systems including systems with large degree of freedom.
It is also possible to extend our method to construct systems
which have different topological feature (i.e., $S^{(N)}(C)$ in 
Eq.~\eqref{eq:SNdef}) from the present one. 
This will be reported in a future publication~\cite{tck112}.
Also, such a recursive construction might be useful to construct
many-body systems with 
the phase anholonomy. Because quantum
circuits allow such a recursive construction in a straightforward
manner, this suggests that quantum circuits offer an interesting
playground for many-body quantum anholonomies.

\acknowledgments
This research was supported by the Japan
Ministry of Education, Culture, Sports, Science and Technology under
the Grant numbers 22540396 and 21540402, and JST, in part.
SWK was supported by the NRF grant funded by the Korea government
(MEST) (No.2009-0087261 and No.2010-0024644).



\end{document}